\documentclass[conference]{IEEEtran}
\usepackage{cite}
\usepackage{amsmath,amssymb,amsfonts}
\usepackage{algorithmic}
\usepackage{graphicx}
\usepackage{textcomp}
\usepackage{xcolor}
\usepackage{bbold}
\usepackage{multirow}
\usepackage{mathtools}
\graphicspath{{figures/}}
\def\BibTeX{{\rm B\kern-.05em{\sc i\kern-.025em b}\kern-.08em
    T\kern-.1667em\lower.7ex\hbox{E}\kern-.125emX}}
\begin{document}
%
\title{{Gen-A}: Generalizing Ambisonics Neural Encoding to Unseen Microphone Arrays}
\author{\IEEEauthorblockN{Mikko Heikkinen}
\IEEEauthorblockA{\textit{Nokia Technologies}\\
Tampere, Finland \\
mikko.heikkinen@nokia.com}
\and
\IEEEauthorblockN{Archontis Politis}
\IEEEauthorblockA{\textit{Tampere University}\\
Tampere, Finland \\
archontis.politis@tuni.fi}
\and
\IEEEauthorblockN{Konstantinos Drossos}
\IEEEauthorblockA{\textit{Nokia Technologies}\\
Espoo, Finland \\
konstantinos.drosos@nokia.com}
\and
\IEEEauthorblockN{Tuomas Virtanen}
\IEEEauthorblockA{\textit{Tampere University}\\
Tampere, Finland \\
tuomas.virtanen@tuni.fi}
}
\maketitle
\begin{abstract}
Using deep neural networks (DNNs) for encoding of microphone array (MA) signals to the Ambisonics spatial audio format can surpass certain limitations of established conventional methods, but existing DNN-based methods need to be trained separately for each MA. This paper proposes a DNN-based method for Ambisonics encoding that can generalize to arbitrary MA geometries unseen during training. The method takes as inputs the MA geometry and MA signals and uses a multi-level encoder consisting of separate paths for geometry and signal data, where geometry features inform the signal encoder at each level. The method is validated in simulated anechoic and reverberant conditions with one and two sources. The results  indicate improvement over conventional encoding across the whole frequency range for dry scenes, while for reverberant scenes the improvement is frequency-dependent.
\end{abstract}
%
\section{Introduction}
\label{sec:introduction}
Ambisonics is a device-independent representation for spatial audio that can be rendered to different reproduction formats using standard solutions \cite{zotter2019ambisonics,olivieri2019scene, pulkki2018parametric}. It is  widely adopted in the industry for various immersive communication and virtual- or extended reality use cases, being one of the transport formats in the 3GPP Immersive Voice and Audio Services \cite{multrus24-AES}, MPEG-H and MPEG-I standard for immersive audio \cite{herre2015mpeg, herre2024mpeg}, and AOM Immersive Audio Model and Format \cite{hwang2023iamf}.

Ambisonics encoding is often implemented as a time-invariant matrix of filters. The filter matrix is obtained by solving the optimal matching problem between directional responses of the microphone array (MA) used for capturing the audio signal and the spherical harmonics in a least squares sense. This approach is well suited for spherical microphone arrays (SMAs) \cite{zotter2019ambisonics, jin2013design, zotkin2017incident, politis2017comparing} and can be extended also to irregular MAs \cite{laborie2003new, zotkin2017incident, jin2013design, politis2017comparing, bastine2022ambisonics}, found, e.g., in mobile phones and head-mounted displays. These methods are model based, taking the MA geometry into account,
and can easily provide their encoder solution to one geometry description or another upon request. However, such methods are limited by the inability of a small number of microphones to capture the Ambisonic components of the sound field (the ambisonic channels) with equal performance over the frequency range of interest. More specifically, they tend to lose energy or over-amplify microphone noise in the first- or higher- order Ambisonic channels at low frequencies, and distort the spherical harmonic (SH) patterns at high frequencies above the MA spatial aliasing limit. The frequency range of these phenomena depend on the size and arrangement of the MA~\cite{rafaely2015fundamentals}.

Recently, signal-dependent parametric methods have been proposed. They estimate spatial parameters from the input signals and then use those parameters to transform the microphone signals to Ambisonic ones in an adaptive manner~\cite{mccormack2022parametric}. They are more flexible in terms of MA geometry and can surpass limitations of signal-independent encoding, but they assume a parametric sound field model to do so. Their encoding performance is dependent on how well the sound scene conforms to this assumed model.

An alternative to the above is non-linear signal-dependent methods that do not assume a simple parameterised sound field model but can learn from data to transform optimally the MA signals to Ambisonic ones.
For example, according to~\cite{gao2022sparse} the
encoding performance can be increased above spatial aliasing frequency for higher-order SMAs using a convolutional neural network.
In~\cite{heikkinen2024neural} is
shown that first order Ambisonics encoding performance for compact irregular MAs can be increased in low frequencies as well as above spatial aliasing frequency using a deep neural network (DNN). However, the above methods need to train the model specifically for a certain MA. Such methods miss the advantage of conventional or parametric solutions to adapt rapidly to a new device or geometry.  
However,~\cite{hsu2023model,hsu2024tunable} propose a DNN-based solution for MA to binaural (not Ambisonics) encoding, with~\cite{hsu2024tunable} employing a spatial coherence input feature that enables their method to offer a level of generalization to unseen MA geometries.

In this work we explore a DNN-based method, employing a neural Ambisonics encoder to achieve improved encoding performance as well as generalization to unseen MA geometries.
The MAs have a fixed number of microphones while the geometry can vary within a range defined by maximum and minimum distances between microphone pairs. The method takes as input MA geometry and MA signals, encodes them through different learning paths, and combines and decodes the learned information through a decoder, predicting the corresponding Ambisonics signals. 
The method is trained on simulated data of sound scenes augmented by a set of MAs and evaluated against a traditional static time-invariant baseline method. Evaluation shows performance surpassing conventional encoding across tested geometries at the problematic low-frequency and high-frequency ranges with few-source dry scenarios. However, we see a drop in performance in reverberant scenarios with baseline outperforming in its optimal encoding range even though the DNN shows benefit at frequencies below and above that range.
%
%
\section{Proposed Method}
\label{sec:methods}
The proposed method takes as an input a set of parameters $\mathbb{\Omega}$, describing the geometry of an MA of $Q$ microphones, and a set of signals $\mathbf{x} = [x_1,...,x_Q]^\mathrm{T}$ from the MA. It then predicts a complex-valued multichannel mixing matrix $\mathbf{E}$ that when multiplied with $\mathbf{x}$ yields the encoded Ambisonic signals $\mathbf{\hat{b}}$.

In more detail, the sound field in Ambisonics is modeled as the SH transform of the continuous directional distribution of plane wave amplitudes expressing the sound scene. Following the definitions in \cite{politis2017comparing}, the Ambisonics signal set up to SH (or Ambisonic) order $N$, is
\begin{equation}
\begin{aligned}
\mathbf{b}(t, f)&=\iint a(t,f,\theta,\phi)\mathbf{y}_N(\theta,\phi)\sin(\phi)d{\theta}d{\phi}.
\end{aligned}
\label{eq:ambisonics_model}
\end{equation}
where $\mathbf{y}_N = [Y_{00},...,Y_{nm},...,Y_{NN}]^\mathrm{T}$ is a vector of real-valued SHs for azimuth $\theta$ and elevation angle $\phi$, indexed by SH order $n$ and degree $m$ up to a maximum encoding order $N$. $\mathbf{b}=[b_{00},...,b_{nm},...,b_{NN}]^\mathrm{T}$ are the $(N+1)^2$ ambisonic signals indexed in the same manner as the SHs and  $a(t, f, \theta,\phi)$ is a plane wave amplitude distribution describing the incident sound field. Assuming time-frequency transformed signals, $t$ is the time and $f$ is the frequency index.

Similarly, the signals captured by an MA characterised by the array transfer functions (ATFs) $\mathbf{h} = [h_1,...,h_Q]^\mathrm{T}$ in the same sound field, are defined as
\begin{equation}
\mathbf{x}(t,f,\mathbb{\Omega})=\iint a(t,f,\theta,\phi) \mathbf{h}(f,\theta,\phi; \mathbb{\Omega})\sin(\phi)d{\theta}d{\phi}
\end{equation}

In this work we focus on MAs of ideal omnidirectional microphones, where ATFs are defined only by the locations of the microphones. We define $\mathbb{\Omega}$ to be the $x, y, z$ coordinates of the microphones. Our aim is to model a mapping function that estimates a $(N+1)^2\times Q$ time- and frequency-dependent encoding matrix $\mathbf{E}(t,f)$ for given $\mathbb{\Omega}$ and $\mathbf{x}$. Multiplying $\mathbf{E}$ and $\mathbf{x}$ yields the estimated Ambisonics encoded signal, ${\mathbf{\hat{b}}}(t, f)$, as
\begin{equation}
{\mathbf{\hat{b}}}(t, f) = \mathbf{E}(t,f) \mathbf{x}(t, f).
\label{eq:estimation_model}
\end{equation}
An optimization criterion is selected to minimize the error between the reference $\mathbf{b}$ and the estimated $\mathbf{\hat{b}}$ Ambisonics signals.
\subsection{DNN Architecture}
\label{ssec:dnn_method}
\begin{figure}
  \centering
  \includegraphics[width=0.99\columnwidth]{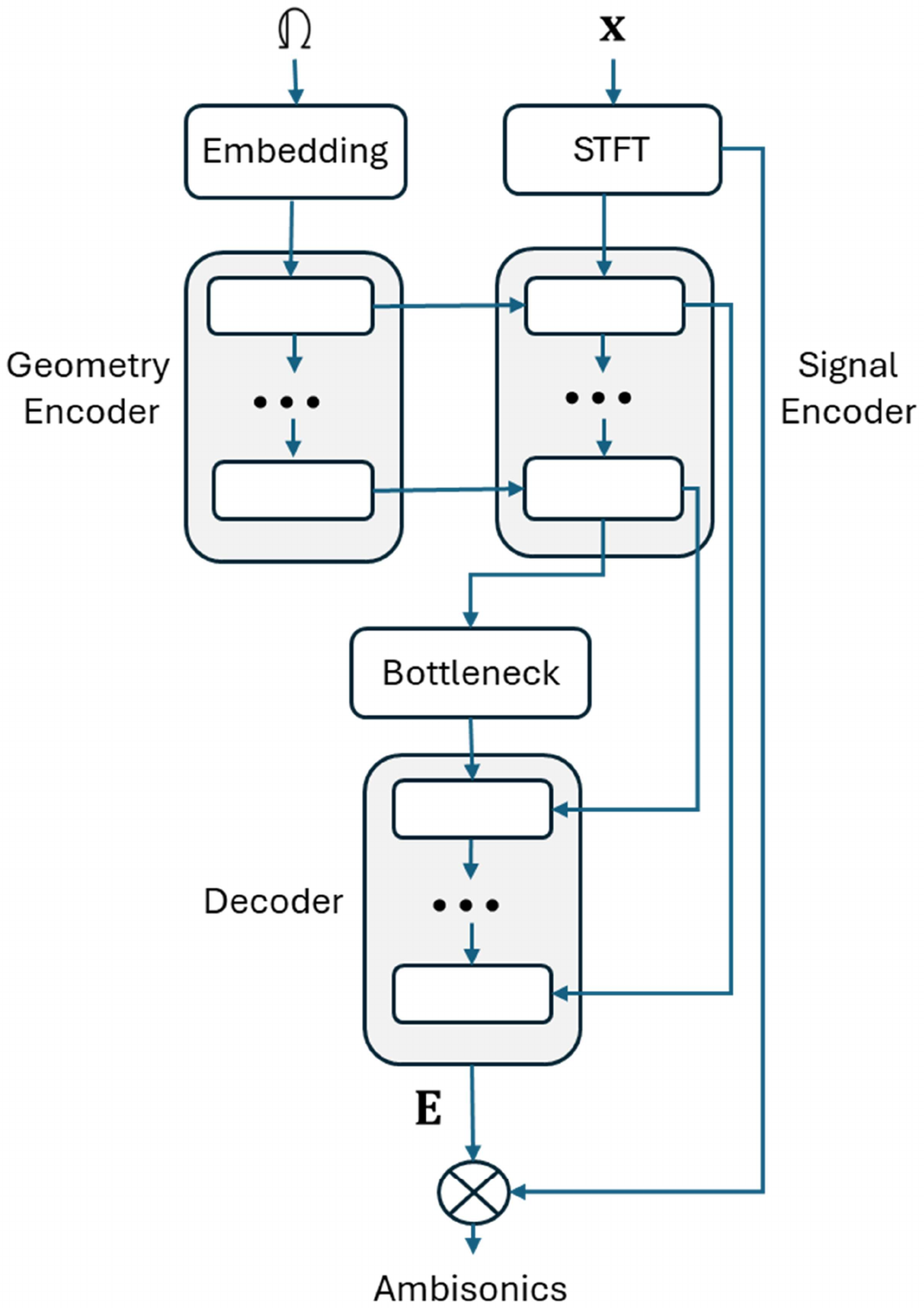}
  \caption{Block diagram of the proposed method, where array geometry ($\mathbb{\Omega}$) and array signals ($\mathbf{x}$) are given as input. $\mathbf{E}$ is the estimated signal- and geometry-dependent complex encoding matrix and $\bigotimes$ is used to denote matrix multiplication along the channel dimension.}
  \label{fig:dnn_overview}
\end{figure}
A block diagram of the proposed DNN
is presented in Figure~\ref{fig:dnn_overview}. The DNN takes as input MA geometry $\mathbb{\Omega}$ and MA signals $\mathbf{x}$. $\mathbf{x}$ is transformed to time-frequency representation using the short-time Fourier transform (STFT). Real and imaginary parts of the complex-valued STFT data are separated into two separate real-valued feature channels to produce input with shape $[2Q, F, T]$, where $F$ and $T$ are frequency and time. $\mathbb{\Omega}$ consists of quantized $x$, $y$, and $z$ MA coordinates which are
projected to an $F$-dimensional space using a learnable mapping from $[Q,3]$ to $[3Q,F]$, then replicated across time to $[3Q,F,T]$.

The DNN predicts a complex-valued mixing matrix $\mathbf{E}$, which is multiplied with the input signals to produce the final Ambisonics signals. The DNN architecture
is based on U-Net~\cite{ronneberger2015u},
characterized by an encoder, bottleneck, and decoder with skip-connection between each encoder and decoder. The proposed method employs separate encoders for the geometry and signal data. Geometry encoder comprises a strided 2D convolution, which reduces
the data size in time and frequency dimensions in the same ratio as the corresponding signal encoder layer. The geometry data is processed using a 2D convolution with 1x1 kernel to match the number of feature channels in the signal encoder data and applied to the corresponding signal encoder layer input data with element-wise multiplication.

The signal encoder consists of two consecutive 2D convolutions with non-linearities followed by channel normalization and dropout. The first convolution of the first encoder layer is a sub-band convolution according to~\cite{kao19_interspeech} to better model the frequency-dependent phase differences caused by sound reaching microphones in the array according to their angle of incidence to the microphone. The second convolution is a strided convolution that reduces the data size in time and frequency dimensions. Each decoder block consists of a 2D transposed convolution, which is concatenated with a skip connection, followed by a dropout layer and two consecutive 2D convolutions with swish activation functions. The bottleneck layer has two consecutive 2D convolutions with swish activations. The skip connections are taken from the signal encoder layers after the dropout layer before max pooling is applied.
%
%
\section{Evaluation}
\label{sec:evaluation}
\subsection{Data}
\label{ssec:data}
Simulated audio scenes were used to train and evaluate the system. The parameters changing between each scene were sound source locations, capturing location, room dimensions, and material properties. The reference Ambisonics signal is simulated in the capturing location with ideal SH encoding at all frequencies. The signals captured by any MA centered at the capturing location must produce the same reference Ambisonics encoding. This allowed us to augment each scene with multiple MAs to produce examples producing the exact same output for different geometry input, which proved to be key factor for our method to learn the given task.

Multichannel room impulse responses (RIRs) were produced for the reference Ambisonics signal and the MA signals using the image-source method with the Pyroomacoustics package~\cite{scheibler2018pyroomacoustics}. Reference Ambisonics RIRs were produced by simulating an MA with ideal Ambisonics directional responses defined by real-valued spherical harmonic coefficients. The MAs were simulated as acoustically transparent with ideal omnidirectional microphones. Thus the set of parameters used for the geometry encoder were the $x$, $y$, $z$ coordinate values of the array microphones.

The ESC-50 dataset \cite{piczak2015esc} containing a wide range of sound event samples was used for the source signal data to make the method robust to any type of signals. Ambisonics and MA signals were obtained by convolving the source signals with the multichannels RIRs. During training the source signal material was selected randomly for each scene, resulting in different audio examples between scenes and training epochs. The levels of source audio material were not altered during processing. No noise or silence was inserted to audio samples. Silence was also not removed from audio samples.

A training dataset with 300 scenes augmented with 1000 MAs were created. The MAs used to augment each scene were randomly sampled from a set of 10000 MAs. MAs had 5 microphones with randomly choosen geometries defined by maximum and minimum distance between microphones as 0.18 m and 0.02 m, respectively. The microphone coordinates were quantized to $24\times24\times24$ grid with length of sides equal to maximum distance between microphones. Each scene was also simulated twice: in a dry setting without room acoustics and in a reverberant setting with a defined room. Room sizes were randomized between $[3, 8]$, $[3,12]$, $[3, 20]$ meters height, width, and depth. Minimum distance of MA to any wall was 1 m, and minimum distance of a sound source to the MA was 2 m. Room materials were selected so that reverberation T60 value was between 0.4 and 0.5 s.

Validation and evaluation datasets with 1000 scenes augmented with 10 MAs each were created. Each MA for each scene was randomly selected and was different from the MAs used in the training dataset. The audio dataset was split with a 80/10/10 split so that no source audio material was shared between the datasets.
\subsection{Employed Hyper-parameters}
A sampling rate of 24 kHz was used. The STFT used an FFT length of 1024 samples with 512 hop size and a Hann window.
The input size was $94\times513,~[T, F]$ for the signal and $5\times3$ for the geometry 
encoder. The quantized coordinates of each microphone in the array were encoded as three integer values from 0 to 24. U-Net input size was zero padded to $96\times560$. Each signal encoder layer halved the time and frequency dimensions of the output of the previous layer. The number of output channels for signal encoder were $\{32, 64, 128, 256\}$. The number of output channels for the geometry encoder was 15 for each layer.
Sub-band convolution was used in 1st signal encoder layer, with sub-band size 1 and kernel size 3x3.
The number of output channels for the bottleneck and decoder layers were
512, and {256, 128, 64, 32}, respectively. Convolution kernel sizes were 3x3 in all layers. Dropout probabilities were 0.25 for signal encoder and 0.5 for decoder layers.

A complex-valued $L_1$-loss was used as the loss function. Training used Adam optimizer, batch size of 32, and learning rate of 2e-4, optimized using the validation data.
\begin{figure*}
  \centering
  \includegraphics[width=\textwidth]{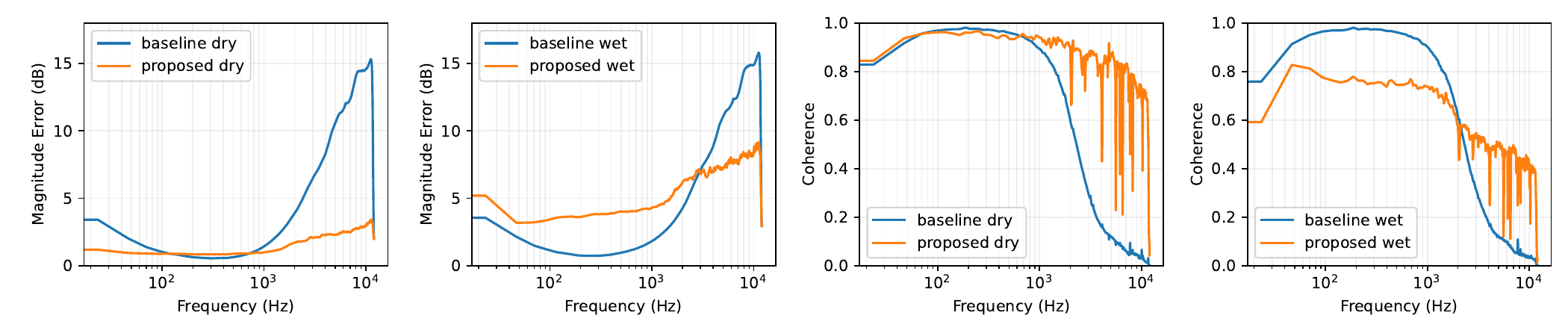}
  \caption{Mean magnitude spectrum error and coherence metrics for proposed and baseline methods in anechoic and reverberant conditions.}
  \label{fig:evaluation_results}
\end{figure*}
\subsection{Baseline}
\label{ssec:baseline}
A conventional static encoding method was selected as baseline based on the review of methods in \cite{politis2017comparing}. The selected method, by Moreau et al. \cite{moreau20063d}, produces a static time-invariant filter matrix by minimizing the least-squares error between ideal and measured Ambisonics responses:
\begin{equation}
\mathbf{E}(f)=\mathbf{Y}^{\mathrm{T}}\mathbf{H}^{\mathrm{H}}(f)\left(\mathbf{H}(f)\mathbf{H}^{\mathrm{H}}(f)+\beta^{2}\mathbf{I}_{\mathrm{Q}}\right)^{-1}.
\end{equation}
$\mathbf{E}$ here is the complex-valued filter matrix that produces the estimated Ambisonics signal after convolving and summing with the array signals. $\mathbf{H}$ is a $Q\times D$ matrix of measured array responses at $D$ directions and $\mathbf{Y}$ is $(N+1)^2\times D$ matrix of SH coefficients for the same directions. $\beta$ is a regularization coefficient that controls maximum amplification in the produced filter matrix. Array responses were simulated and evaluated in a uniformly distributed grid of $D=1008$ directions. Filter amplification was not limited to have comparable results with proposed method. 
\subsection{Metrics}
\label{ssec:metrics}
Spectral errors were evaluated with a mean magnitude spectrum error across individual channels:
\begin{equation}
S(f)=\frac{1}{TC}\sum_{c=1}^{C}\sum_{t=1}^{T}\left|20\cdot \log_{10}\left( \frac{|b_{t,f,c}|}{|\hat{b}_{t,f,c}|}\right)\right|
\label{eq:magnitude_spectrum_err}
\end{equation}
The spatial quality of the encoding was evaluated with a magnitude squared coherence metric: 
\begin{equation}
C(f) = \frac{1}{C}\sum_{c=1}^C\frac{|\sum_{t=1}^{T} b^*_{t,f,c} \hat{b}_{t,f,c} |^2 }{ \sum_{t'=1}^T |b_{t',f,c}|^2 \sum_{t''=1}^T |\hat{b}_{t'',f,c}|^2 }
\label{eq:coherence}
\end{equation}
where $b^{\ast}$ is the complex conjugate of the reference Ambisonics signal. 

Overall signal degradation of the estimated encoding was evaluated with scale-invariant signal-to-noise ratio (SI-SNR) as defined in \cite{luo2018tasnet}.
\subsection{Results}
\label{ssec:results}
Figure~\ref{fig:evaluation_results} compares the results between the proposed and baseline method. Evaluation with anechoic data (labeled "dry"), shows that the proposed method far exceeds the performance of the baseline method for one and two sources. Both the magnitude error and coherence shows that the baseline method performance quickly degrades above the spatial aliasing frequency, while the proposed method operates with an acceptable error across the entire frequency range. Evaluation on reverberant conditions (labeled "wet") shows that the proposed method is sensitive to reverberation effects such as correlated early echoes and diffuse late reverberation and in those conditions it does not meet the performance of the baseline method. However, the performance of proposed method is more consistent across the full frequency range with less degradation above the spatial aliasing frequency.

Table~\ref{tab:si-snr} further details the difference in operation between methods in various conditions. The proposed method shows very good performance in anechoic conditions but with considerable difference in SI-SNR values between conditions with only one sound source and two active sound sources. Performance in reverberant conditions is overall much poorer and also shows degradation in results when adding a second active sound source. The baseline method shows more consistent performance across various conditions, which is expected for a signal-independent encoder.
Unlike the baseline method, the proposed method has no mechanism to limit the maximum gain applied by the encoding matrix, which can potentially lead to unpleasant amplification of microphone noise. This aspect has not been evaluated yet and is left for follow up work. Also, the current evaluation consists of random sampling of sound scenes and arrays, so it does not give insight on how consistently an individual array behaves under different conditions. A detailed analysis of the method on array-specific performance for a range of conditions is left for follow up work.

\begin{table}
\caption{Metrics averaged over all frequencies}
\centering
\begin{tabular}{l|l|ll|ll|}
\cline{2-6}
                                                       & \multicolumn{1}{c|}{\multirow{2}{*}{}} & \multicolumn{2}{c|}{Single Source}                 & \multicolumn{2}{c|}{Dual Source}                   \\ \cline{3-6} 
                                                       & \multicolumn{1}{c|}{}                  & \multicolumn{1}{l|}{Dry}           & Wet           & \multicolumn{1}{l|}{Dry}           & Wet           \\ \hline
\multicolumn{1}{|l|}{\multirow{2}{*}{SI-SNR}}          & \multicolumn{1}{c|}{Baseline}          & \multicolumn{1}{c|}{5.2}           & \textbf{5.6}  & \multicolumn{1}{c|}{5.7}           & \textbf{6.3}  \\ \cline{2-6} 
\multicolumn{1}{|l|}{}                                 & \multicolumn{1}{c|}{Proposed}          & \multicolumn{1}{c|}{\textbf{23.2}} & 4.3           & \multicolumn{1}{c|}{\textbf{14.3}} & 3.0           \\ \hline
\multicolumn{1}{|l|}{\multirow{2}{*}{Coherence}}       & Baseline                               & \multicolumn{1}{l|}{0.25}          & 0.25          & \multicolumn{1}{l|}{0.24}          & 0.24          \\ \cline{2-6} 
\multicolumn{1}{|l|}{}                                 & Proposed                               & \multicolumn{1}{l|}{\textbf{0.93}} & \textbf{0.57} & \multicolumn{1}{l|}{\textbf{0.69}} & \textbf{0.44} \\ \hline
\multicolumn{1}{|l|}{\multirow{2}{*}{Magnitude Error}} & Baseline                               & \multicolumn{1}{l|}{9.0}           & 9.5           & \multicolumn{1}{l|}{10.6}          & 11.0          \\ \cline{2-6} 
\multicolumn{1}{|l|}{}                                 & Proposed                               & \multicolumn{1}{l|}{\textbf{1.2}}  & \textbf{5.9}  & \multicolumn{1}{l|}{\textbf{3.6}}  & \textbf{8.2}  \\ \hline
\end{tabular}
\label{tab:si-snr}
\end{table}
%
%
\section{Conclusions}
\label{sec:conclusion}
In this work, we proposed a DNN-based Ambisonic encoding method for microphone array signals which takes geometry and signal data as input and generalizes to unseen microphone arrays. The method is based on a U-Net-like architecture with separate encoders for geometry and signal data and a method to apply data from geometry encoder layers to signal encoder layers. The method was trained with simulated sound scenes and augmentations of each scene with various microphone arrays to learn the effect of array geometry on the encoding. Based on objective encoding metrics, the evaluation shows that the method learns to generalize and, when compared against a conventional Ambisonic encoder, it performs very well in acoustically dry conditions, but with a performance drop in reverberant ones.

Topics for further research include extending the method to use array data with realistic directional magnitude and phase responses in training and evaluation, as well as improving the DNN structure and training data to improve performance in realistic conditions containing reverberation, multiple sound sources, diffuse background sound, and microphone noise. Finally, while the evaluation metrics used in this study are relevant for use cases where Ambisonics is the transport format, listening tests will be conducted to assess the actual perceptual performance of the encoder compared to other methods.
\clearpage
%
\bibliographystyle{IEEEtran}
\bibliography{IEEEabrv,refs}
\end{document}